\documentclass[prl,twocolumn,twoside,preprintnumbers,superscriptaddress,nofootinbib]{revtex4}

\usepackage{amsmath,amssymb,bm,graphicx,bbold,epsf,colordvi}
\usepackage[a4paper,text={16.8cm,22.4cm}]{geometry}
\usepackage{lipsum}
\allowdisplaybreaks 
\newcommand\slurp[1]{#1}
\newcommand\sslurp[2]{#1{#2}}
{\catcode`/=\active \expandafter}%
\slurp{\newcommand/}{/\penalty1000\hskip0pt\relax}
{\catcode`:=\active \expandafter}%
\slurp{\newcommand:}{:\penalty1000\hskip0pt\relax}

\newcommand\addspace{\ifcat\nextchar a\spacefactor999. \else.\fi}
{\catcode`\.=\active \expandafter}%
\slurp{\newcommand.}{\futurelet\nextchar\addspace}

\usepackage[unicode]{hyperref} 
\ifx\href\undefined\def\href#1{}\fi
\ifx\texorpdfstring\undefined\def\texorpdfstring#1#2{#1}\fi

\newcommand\myslash{/} \newcommand\mycolon{:}
\newcommand\doi{{\catcode`/=\active \catcode`:=\active \expandafter}\sslurp\realdoi}
{\catcode`/=\active \catcode`:=\active \expandafter}%
\slurp{\newcommand\realdoi[1]{{\let/=\myslash \let:=\mycolon
                               \edef\raw{{http://dx.doi.org/#1}}\expandafter}%
                               \expandafter\href\raw{doi:#1}}}
\newcommand\eprint[2]{{\escapechar-1%
                       \edef\a{\expandafter\string\csname arXiv\endcsname}%
                       \edef\b{\expandafter\string\csname #1\endcsname}%
                       \edef\c{\expandafter\string\csname #2\endcsname}%
                       \edef\d{\noexpand\href{http://arXiv.org/abs/\c}}%
                       \ifx\a\b\expandafter\d\fi{\tt #1:#2}}}
\newcommand{\be}{\begin{equation}}
\newcommand{\ee}{\end{equation}}

\def\d{{\rm d}}
\def\OMIT#1{{}}

\newcommand{\e}{\mathrm{e}}

\newcommand{\eq}[1]{Eq.~\eqref{#1}}

\newcommand{\vc}[1]{\boldsymbol{#1}}

\def\SCETG{${\rm SCET}_{\rm G}\,$}

\begin{document}
%%%%%%%%%%%%%%%%%%%%%%%%%%%%%%%%%

\preprint{\vbox{\hbox{ACFI-T14-10}}}

\title{Jet quenching phenomenology  \\ from soft-collinear effective theory with Glauber gluons }
\author{Zhong-Bo Kang}
\affiliation{Theoretical Division, Los Alamos National Laboratory 
MS B283, Los Alamos, NM 87545 USA}%
\author{Robin Lashof-Regas}
\affiliation{Theoretical Division, Los Alamos National Laboratory 
MS B283, Los Alamos, NM 87545 USA}
\affiliation{ Department of Physics,
University of California,
Santa Barbara, CA 93106 USA}
\author{Grigory Ovanesyan}
\affiliation{Physics Department, University of Massachusetts Amherst, Amherst, MA 01003, USA}%
\author{Philip Saad}
\affiliation{Theoretical Division, Los Alamos National Laboratory 
MS B283, Los Alamos, NM 87545 USA}
\affiliation{ Department of Physics,
University of California,
Santa Barbara, CA 93106 USA}
\author{Ivan Vitev}
\affiliation{Theoretical Division, Los Alamos National Laboratory 
MS B283, Los Alamos, NM 87545 USA}%

\date{May 11, 2014}
%%%%%%%%%%%%%%%%%%%%%%%%%%%%%%%%%%
%%%%%%%%%%%%%%%%%%%%%%%%%%%%%%%%%%
\begin{abstract}
We present the first application of a recently-developed effective theory of jet propagation in matter \text{\SCETG}
to inclusive hadron suppression in nucleus-nucleus collisions at the LHC. \text{\SCETG}-based splitting kernels 
allow us to go beyond the traditional  energy loss approximation and unify  the treatment of vacuum and medium-induced 
parton showers. In the soft gluon emission limit, we establish a simple analytic relation between the QCD evolution and
energy loss approaches to jet quenching. We quantify the uncertainties associated with the implementation of the in-medim 
modification of hadron production cross sections  and show that the coupling between the jet and the medium can be 
constrained with better than 10\% accuracy. 
     
 \end{abstract}
%
%\pacs{10.10.Gh, 12.60.Jv, 14.40.Nd, 14.70.Fm, 14.80.Ly}

\maketitle

Suppression of the production cross section for high transverse momentum particles and jets in ultra-relativistic
collisions of heavy nuclei, commonly referred to as jet quenching~\cite{Wang:1991xy},  is one of the most-important signatures 
of quark-gluon plasma (QGP) formation in such reactions and a quantitative probe of its properties. This phenomenon has been
established experimentally  at the Relativistic Heavy Ion Collider (RHIC)~\cite{Adler:2003qi,Adams:2003kv} and 
the Large Hadron Collider (LHC) \cite{Aamodt:2010jd,CMS:2012aa,Aad:2012vca}.  It was   understood 
theoretically  in a framework  based on perturbative QCD calculations of parton propagation and 
energy loss in the QGP~\cite{Gyulassy:2003mc}.

More recently, progress has been made on formulating and applying effective theories of QCD, suitable for calculations of jet properties in hot and dense strongly-interacting matter. The well-established soft-collinear effective theory (SCET)~\cite{Bauer:2000ew,Bauer:2000yr} has been extended to include the interactions with the medium quasiparticles via a transverse $t$-channel momentum exchange. The resulting soft-collinear effective theory with Glauber gluons (\text{\SCETG})~\cite{Idilbi:2008vm,Ovanesyan:2011xy} has been used to calculate all 
${\cal O}(\alpha_s)$ $1\rightarrow 2$ medium-induced splitting kernels~\cite{Ovanesyan:2011kn}  and  study  ${\cal O}(\alpha_s)$ effects on the in-medium parton shower~\cite{Fickinger:2013xwa}.  The power counting of \text{\SCETG} correctly captures the 
behavior of the in-medium branchings when the  lightcone momentum  fraction  $x=Q^+/p^+$ of the {\it emitted} parton becomes large ($x \rightarrow 1$).  These  large-$x$ corrections are absent in  traditional  energy loss calculations.

A critical step in improving the jet quenching phenomenology is to understand the implication of the finite-$x$ corrections. Their implementation
requires new theoretical methods, since in the large momentum fraction  limit the leading parton can change flavor and the splitting process cannot be 
interpreted as  energy loss. A natural language to capture this physics is that of  the well-known  DGLAP evolution 
equations~\cite{Altarelli:1977zs}.  As a  first application of the \text{\SCETG}  medium-induced splitting kernels, we revisit the evaluation of the 
nuclear modification factor $R_{AA}$ for inclusive hadron production at high transverse momentum $p_T$ (and rapidity $y$), defined as:  
\begin{eqnarray} 
&&R_{AA} (p_T)  = 
\frac{ d \sigma^h_{AA}/dyd^2 p_T }
{   \langle N_{ \rm coll}\rangle    d \sigma^h _{pp}  / dyd^2 p_T } \, , \label{raa-analyt} 
\end{eqnarray} 
which continues to attract strong theoretical interest~\cite{Betz:2014cza,Djordjevic:2013xoa}. 
We consider  central lead-lead (Pb+Pb) reactions  at $\sqrt{s_{NN}}=2.76$~TeV at  the LHC as an example. 
In Eq.~(\ref{raa-analyt})  ${\langle N_{ \rm coll}\rangle}$ is the average number of binary nucleon-nucleon collisions.
DGLAP evolution equations have been used to address hadron production in semi-inclusive deep
inelastic scattering with initial conditions obtained using an energy loss approach~\cite{Wang:2009qb,Chang:2014fba}.

In the presence of a QGP, all parton splitting kernels are a direct sum of the universal vacuum part  and a  medium-dependent 
component, which has  been calculated in Ref.~\cite{Ovanesyan:2011kn}. Those are real emission graphs in the DGLAP language. 
The splitting functions are related to the medium-induced splitting kernels as follows:
\begin{eqnarray}
P^{\text{real}}_{i}(x,\vc{Q}_{\perp};\alpha)&=&\frac{2\pi^2}{\alpha_s}\,\vc{Q}_{\perp}^2\,\frac{\d N_i (x, \vc{Q}_{\perp};\alpha)}{\d x\,\d^2\vc{Q}_{\perp}}\nonumber\\
&=& P^{\text{vac}}_i(x)\,g_i(x,\vc{Q}_{\perp};\alpha)\, .
\end{eqnarray}
The equation above explicitly indicates that, unlike the vacuum case where the splitting function only depends on $x$, the medium-induced splitting function also depends on $\vc{Q}_{\perp}$ and the properties of the QGP $\alpha$.   We relate the temperature and density of the gluon-dominated plasma to the measured charged particle rapidity density~\cite{Gyulassy:2003mc}. The position and time dependence of the Debye screening scale $m_D$ and the quark and gluon scattering lengths,   necessary to evaluate  $P^{\text{real}}_{i}(x,\vc{Q}_{\perp};\alpha)$,  are obtained using an optical Glauber model for the collision geometry and a Bjorken expansion ansatz. The coupling $g$ between the jet and the medium is a free parameter in the calculation.  

Special attention has to be paid to the gluon splitting function because it diverges for both $x\rightarrow 0$ and $x\rightarrow 1$. The first divergence is regulated with a plus function prescription, while the second divergence need not be regulated owing to  the form of the evolution equations:
\begin{eqnarray}
&&P_{q\rightarrow qg}(x)=\left[P^{\text{real}}_{q\rightarrow qg}(x)\right]_++A\,\delta(x)\, ,\\
&&P_{g\rightarrow gg}(x)=2C_A\Bigg\{\left[\left(\frac{1-2x}{x}+x(1-x)\right)g_2\left(x\right)\right]_+\nonumber\\
&&\qquad\qquad\qquad\qquad+\frac{g_2\left(x\right)}{1-x}\Bigg\}+B\,\delta(x)\, ,\\
&&P_{g\rightarrow q\bar{q}}(x)=P^{\text{real}}_{g\rightarrow q\bar{q}}(x)\, , \quad  P_{q\rightarrow gq}(x)=P^{\text{real}}_{q\rightarrow gq}(x)\, .
\end{eqnarray}
In the equations above we have suppressed the explicit  $\vc{Q}_{\perp}$ and $\alpha$ dependence for simplicity. 
The virtual pieces of the splitting functions can be extracted from  flavor and momentum sum rules in complete analogy to the vacuum case: 
\begin{eqnarray}
&&A = 0\, , \\
&&B =\int_0^1\d x'\Bigg\{-2n_f(1-x')P_{g\rightarrow q\bar{q}}(x')\nonumber\\
&&+2C_A\Bigg[x'\left(\frac{1-2x'}{x'}+x'(1-x')\right)-1\Bigg]g_2\left(x'\right)\Bigg\}\, . \quad
\end{eqnarray}
The DGLAP evolution equations for the fragmentation functions (FFs)  read:
%\end{widetext}
\begin{eqnarray}
&&\frac{\d D_q(z,Q)}{\d \ln Q}=\frac{\alpha_s(Q^2)}{\pi}\int_{z}^1 \frac{\d z'}{z'}\Big[P_{q\rightarrow qg}(z')D_q\left(\frac{z}{z'},Q\right)\nonumber\\
&&\qquad\qquad\qquad+P_{q\rightarrow gq}(z')D_g\left(\frac{z}{z'},Q\right)\Big]\, ,\label{eq:AP10}\\
&&\frac{\d D_g(z,Q)}{\d \ln Q}=\frac{\alpha_s(Q^2)}{\pi}\int_{z}^1 \frac{\d z'}{z'}\Bigg[P_{g\rightarrow gg}(z')D_g\left(\frac{z}{z'},Q\right)\nonumber\\
&&+P_{g\rightarrow q\bar{q}}(z')\sum_{q}\left(D_q\left(\frac{z}{z'},Q\right)+D_{\bar{q}}\left(\frac{z}{z'},Q\right)\right)\Bigg]\, ,\label{eq:AP30}
\end{eqnarray}
where  $z \equiv 1-x $ in the splitting functions  and $Q \equiv |\vc{Q}_{\perp}|$. The equation for the evolution of the 
anti-quark FF can be found from quark equation by substituting everywhere $D_q\rightarrow D_{\bar{q}}$.

QCD evolution and the energy loss approach represent two very different implementations of jet quenching. It is critical to establish 
this connection between them in light of the fact that energy loss phenomenology has been very 
successful~\cite{Gyulassy:2003mc,Djordjevic:2013xoa,Betz:2014cza}.  This can be achieved {\em only} in the soft gluon 
bremsstrahlung  limit,  where the two diagonal splitting functions $P_{q\rightarrow qg}$ and $P_{g\rightarrow gg}$ survive. 
 Up to $({2\pi^2}/{\alpha_s})\,\vc{Q}_{\perp}^2$, these are
 the Gyulassy-Levai-Vitev (GLV) double differential medium-induced gluon number distributions to first order in opacity~\cite{Gyulassy:2000fs}.
 There is no flavor mixing, and the entire 
 branching is given by a plus function. The DGLAP evolution equations decouple and reduce to:
\begin{eqnarray}
\frac{\d D(z,Q)}{\d\ln Q}=\frac{\alpha_s}{\pi}\int_z^{1}\frac{\d z'}{z'}\,\left[P(z',Q)\right]_+D\left(\frac{z}{z'},Q\right) \,. \qquad
\label{eq:evolutionfull}
\end{eqnarray}
Because the fragmentation functions $D(z)$ are typically steeply falling with increasing $z = p_T^{\rm hadron}/ p_T^{\rm parton}$, the main 
contribution in \eq{eq:evolutionfull}  
comes predominantly from $z'\approx 1$. We expand the integrand in this limit,  keeping the first derivative terms, and 
approximate the steepness of the fragmentation function with its unperturbed vacuum value:
\begin{eqnarray}
n(z)=-{\d \ln D^{\text{vac}}(z)}/{\d\ln z}\,.
\end{eqnarray}
The analytical solution to the Eq.~(\ref{eq:evolutionfull}) reads: 
\begin{eqnarray}
&&D^{\text{med}}(z,Q)\approx \e^{-(n(z)-1)\left\langle\frac{\Delta E}{E}\right\rangle_z-\left\langle N_g\right\rangle_z}D^{\rm vac}(z,Q)\,, \qquad\label{eq:masterconnect}
\end{eqnarray}
and shows explicitly that the vacuum evolution and the medium-induced evolution factorize.
We have used the following definitions in the above formula:
\begin{eqnarray}
&&\left\langle\frac{\Delta E}{E}\right\rangle_z=\int_0^{1-z}\d x \,x\,\frac{\d N}{\d x}(x) \xrightarrow{z \to 0}\left\langle\frac{\Delta E}{E}\right\rangle\, , \qquad \\
&&\left\langle N_g\right\rangle_z=\int_{1-z}^1\d x\,\frac{\d N}{\d x}(x)\xrightarrow{z\to1}\left\langle N_g\right\rangle\,,
\end{eqnarray}
where $xdN/dx$ is the medium-induced gluon intensity distribution~\cite{Gyulassy:2000fs}.
Note, that we have made the choice to put all the in-medium effects into the DGLAP evolution. The analytic 
formula in \eq{eq:masterconnect} gives us for the first time an insight into the deep connections between the evolution  
and energy loss approaches to  jet quenching. Over most of the $z$ range the suppression of the FFs is dominated by the 
the fractional energy loss, amplified by the steepness of  $D(z)$. Near threshold ($z=1$) the modification is determined by the probability
{\em not} to emit a gluon, $\exp(-\langle N_g \rangle )$. Conversely, solving  Eqs. (\ref{eq:AP10}) and (\ref{eq:AP30}) numerically
allows us to unify the treatment of the vacuum and medium-induced parton showers.

We now  turn to the numerical  comparison between the medium-modified evolution approach to  jet quenching and the traditional 
energy loss  formalism. We elect to include all QGP effects in the fragmentation functions, such that the invariant inclusive hadron 
production cross section reads:
\begin{eqnarray}
 \frac{1}{\langle N_{ \rm coll}\rangle } \frac{d \sigma^{h}_{AA}}{dyd^2p_T} &=&   \sum_c \int_{z_{\min} }^1 dz \; 
\frac{d \sigma^{c} (p_c = p_T/z) }{dyd^2p_{T_c}} \nonumber \\
&& \times \frac{1}{z^2} D^{\rm med/quench}_{c}(z) \, .
\label{hspectrum}
\end{eqnarray}
Here, $c=\{q, \bar{q},g\}$ and we choose the factorization, fragmentation and renormalization scales  $Q=p_{T_c}$, 
and  ${d \sigma^{c} }/{dyd^2p_{T_c}} $ is the unmodified hard parton  production cross section.

Should an energy loss approach be adopted,  it is  important to realize that  the soft gluon emission limit 
must be consistently implemented. If the  fractional energy loss  becomes significant, it is carried away through 
multiple gluon bremsstrahlung. In the independent Poisson gluon emission  limit, we  can construct the probability density  
 $P_c(\epsilon)$ of this  fractional energy loss   $\epsilon  = \sum_{i } \omega_i /E \approx   \sum_{i } Q^+_i/p^+$,
 such that: 
 \begin{eqnarray} 
&& \int_0^1 d \epsilon \; P( \epsilon ) = 1\;, \quad 
 \int_0^1 d \epsilon \; \epsilon \, P( \epsilon ) = 
\left\langle  \frac{ \Delta E }{ E } \right\rangle \;. \quad 
\label{meane} 
\end{eqnarray}
A more detailed discussion is given in~\cite{Gyulassy:2003mc}.  If a parton loses this energy fraction $\epsilon$ during its 
propagation  in the QGP to escape with momentum $p^{\rm quench }_{T_c}$, immediately 
after the hard collision  $p_{T_c} = p^{\rm quench}_{T_c} /  ( 1 - \epsilon )$. Noting the additional Jacobian $|d p_{T_c}^{\rm quench} 
/ d p_{T_c}| = ( 1- \epsilon) $, the kinematic modification to the FFs due to energy loss is:
\begin{eqnarray}
&&D_{c}^{\rm quench}(z) =     \int_0^{1-z} \, d \epsilon\,   \frac{ P_c(\epsilon)  }{(1-\epsilon)}
D_{c}\left(\frac{z}{1-\epsilon}\right)  \, , \quad
\label{quenchD}
\end{eqnarray}
and can be directly implemented in Eq.~(\ref{hspectrum}).

\begin{figure}[!t]
\begin{center}
\includegraphics[width=8cm]{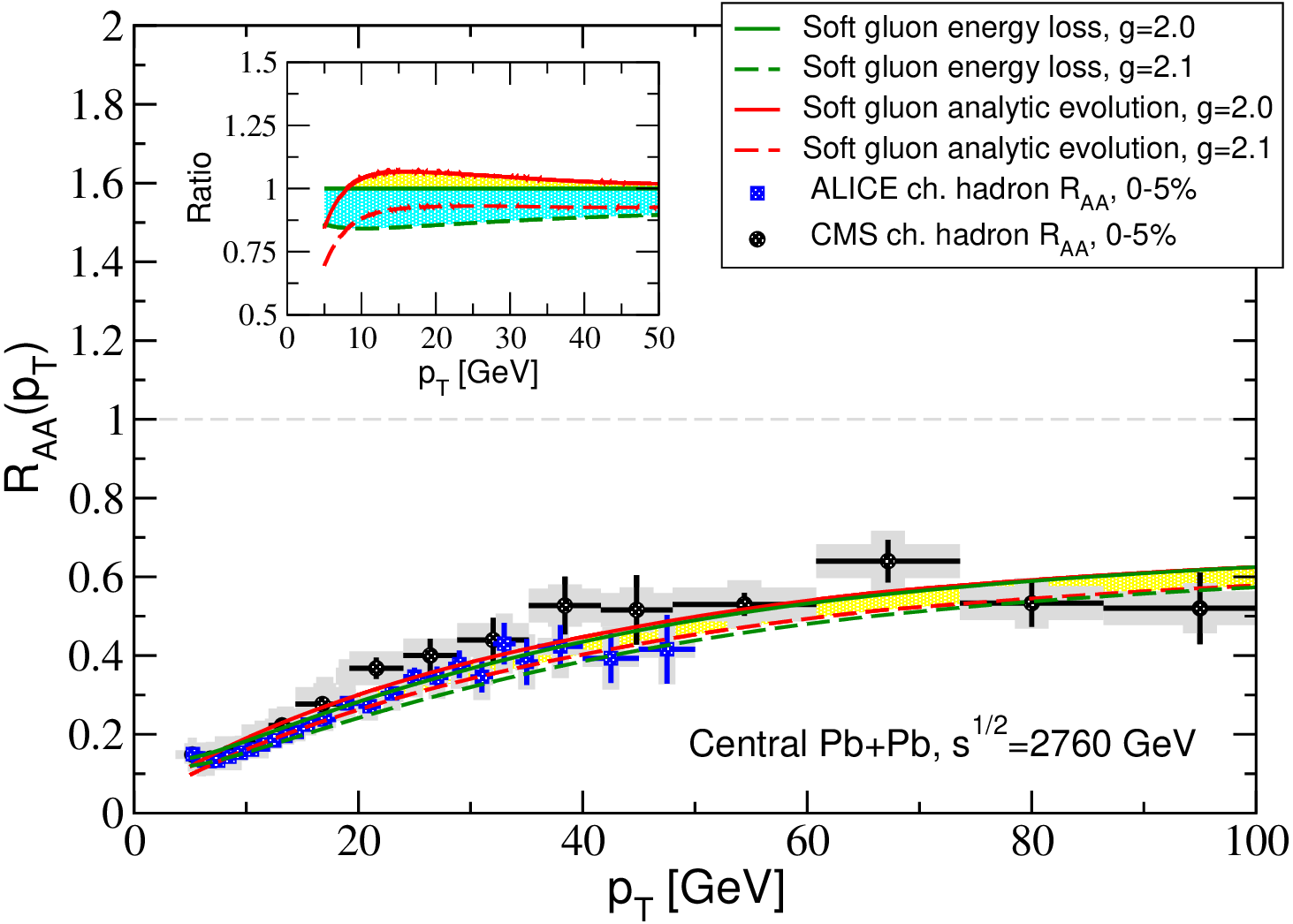}
\vspace*{-6mm}
\caption{Nuclear modification factor comparison between the traditional energy loss approach (cyan band) and the analytic solution to QCD evolution in the soft gluon limit (yellow band). The upper and lower edges of the bands correspond to couplings between the jet and the medium $g=2.0$ and $g=2.1$, respectively.
The insets show the ratios of different $R_{AA}$ curves.  Data is form ALICE an CMS. 
}\label{fig:evolutionRAA1}
\vspace*{-9mm}
\end{center}
\end{figure}

In Figure~\ref{fig:evolutionRAA1} we present our calculations of the nuclear modification factor $R_{AA}$ in the limit of soft gluon bremsstrahlung. 
Results are obtained from the parton energy loss  approach (cyan band) and by using the analytic solution to the in-medium evolution given in 
Eq.~(\ref{eq:masterconnect}) (yellow band). The upper edge of the uncertainty  bands (solid lines) corresponds to a coupling between the jet and the 
medium $g=2.0$ and the lower edge (dashed lines) corresponds to $g=2.1$. The results of the two calculations are remarkably similar and
both reproduce well the suppression of inclusive charged hadron production in 0-10\% central Pb+Pb collisions at the LHC 
measured by ALICE~\cite{Aamodt:2010jd} and CMS~\cite{CMS:2012aa}.  In both approaches the coupling $g$ between the jet and the
medium can be constrained with an accuracy of 5\% and the transport properties of the medium, which scale as $g^4$, can be extracted with 
20\% uncertainty.  The inset shows the ratio for the different $R_{AA}$ curves relative to the $g=2.0$ energy loss  result. We observe from this 
inset  that the only difference between  the two approaches is a small variation in the shape of the nuclear modification ratio 
as a function of $p_T$. At any fixed transverse momentum  the difference in the predicted magnitude of 
jet quenching  can be absorbed in less than 2\% change of the coupling $g$ between the jet and the medium.

\begin{figure}[!t]
\begin{center}
\includegraphics[width=8cm]{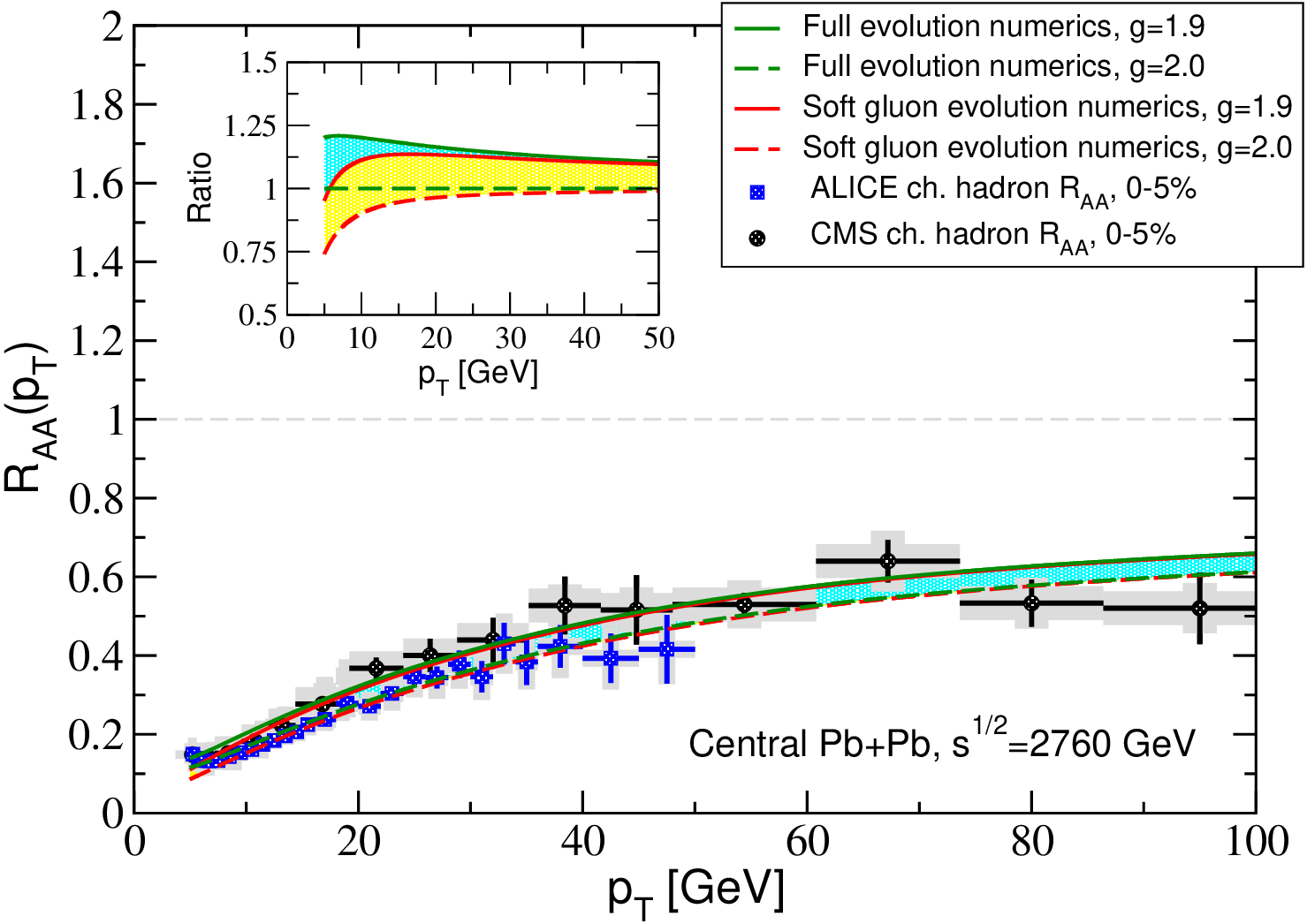}
\vspace*{-5mm}
\caption{ Comparison between $R_{AA}$  obtained with in-medium numerically evolved fragmentation functions using the full splitting kernels 
 (cyan band) and their soft gluon limit (yellow band) to ALICE and CMS data. The upper and lower edges of the bands correspond to $g=1.9$ and $g=2.0$, respectively.
}\label{fig:evolutionRAA2}
\vspace*{-10mm}
\end{center}
\end{figure}

In Figure \ref{fig:evolutionRAA2}  we show $R_{AA}$s obtained with medium-modified FFs 
that are numerical solutions to the DGLAP evolution equations,  Eqs.~(\ref{eq:AP10}), (\ref{eq:AP30}), with  full medium-induced  
splitting kernels~\cite{Ovanesyan:2011kn} (cyan band) and their small-$x$ energy loss limit~\cite{Gyulassy:2000er} (yellow band).
In this figure, the uncertainty bands correspond to $g= 1.9 - 2.0$. The difference between the small-$x$ and full evolution is only 
noticeable below $p_T=20$~GeV, as can be seen from the inset. At small and intermediate transverse momenta 
the solution to the DGLAP equations beyond the soft gluon limit yields a slightly better agreement between theory and experiment.

\begin{figure}[!t]
\begin{center}
\includegraphics[width=7cm,height=8cm]{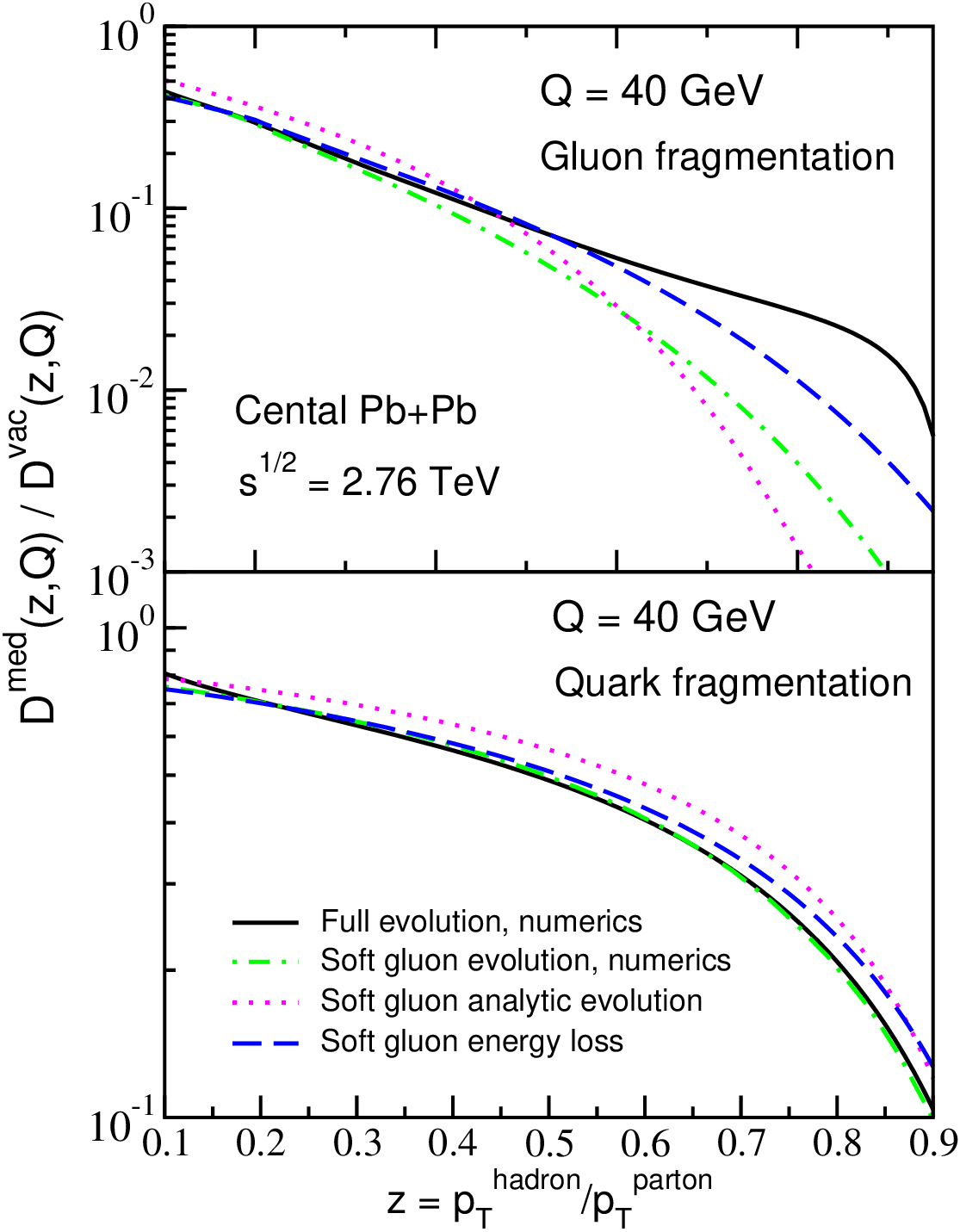}
\caption{The modification of the fragmentation functions for gluons (top panel) and quarks (bottom panel) are shown
for $Q=40$~GeV and central Pb+Pb collisions at the LHC, using four different methods to compute the in-medium 
modification with $g=2.0$.} 
\label{fig:evolutionFF1}
\end{center}
\end{figure}

To understand the numerical results, we further scrutinize the in-medium modification of  FFs in Figure~\ref{fig:evolutionFF1}  
for  40~GeV quarks and gluons, respectively. As a function of the hadron-to-parton transverse momentum fraction $z$, the differences between 
the various methods of computing this modification can 
be much more pronounced than in $R_{AA}$. This is especially  true for gluon fragmentation at large $z$. The observed hadron 
production cross section, however, samples a wide range of momentum fractions and in the presence of a QGP is biased toward lower
values of   $z$. Furthermore, the quark contribution is enhanced since $D^{\rm med}_q(z)$ is much less 
suppressed  than $D^{\rm med}_g(z)$.

To summarize, we presented results for the suppression of inclusive hadron production in Pb+Pb reactions at the LHC  based upon QCD factorization and DGLAP evolution with \text{\SCETG}-based medium-induced splitting kernels. This method allows us to unify the treatment of vacuum and medium-induced parton showers. In the soft gluon bremsstrahlung limit, we demonstrated the connection between this new approach and the traditional energy loss-based jet quenching phenomenology. Numerically, the agreement between the two methods is quite remarkable and they give a very good description of the experimentally measured $R_{AA}$ by ALICE and CMS. We find that the
coupling between the jet and the medium can be constrained with better than 10\% accuracy when the uncertainties that arise from 
the choice of method and the fit to the data are combined. In the future,  it will be interesting to investigate whether better differentiation between
the QCD evolution and energy loss approaches can be achieved using parton flavor separation 
techniques~\cite{Ellis:2010rwa,Gallicchio:2011xq}.

Acknowledgments: This work is supported by DOE Office of Science and in part by the LDRD program at LANL.   

\bibliographystyle{h-physrev}
\bibliography{bibliography}

\begin{thebibliography}{10}

\bibitem{Wang:1991xy}
X.-N. Wang and M.~Gyulassy,
\newblock Phys.Rev.Lett. {\bf 68}, 1480 (1992).

\bibitem{Adler:2003qi}
PHENIX Collaboration, S.~Adler {\em et~al.},
\newblock Phys.Rev.Lett. {\bf 91}, 072301 (2003), nucl-ex/0304022.

\bibitem{Adams:2003kv}
STAR Collaboration, J.~Adams {\em et~al.},
\newblock Phys.Rev.Lett. {\bf 91}, 172302 (2003), nucl-ex/0305015.

\bibitem{Aamodt:2010jd}
ALICE Collaboration, K.~Aamodt {\em et~al.},
\newblock Phys.Lett. {\bf B696}, 30 (2011), 1012.1004.

\bibitem{CMS:2012aa}
CMS Collaboration, S.~Chatrchyan {\em et~al.},
\newblock Eur.Phys.J. {\bf C72}, 1945 (2012), 1202.2554.

\bibitem{Aad:2012vca}
ATLAS Collaboration, G.~Aad {\em et~al.},
\newblock Phys.Lett. {\bf B719}, 220 (2013), 1208.1967.

\bibitem{Gyulassy:2003mc}
M.~Gyulassy, I.~Vitev, X.-N. Wang, and B.-W. Zhang,
\newblock (2003), nucl-th/0302077.

\bibitem{Bauer:2000ew}
C.~W. Bauer, S.~Fleming, and M.~E. Luke,
\newblock Phys.Rev. {\bf D63}, 014006 (2000), hep-ph/0005275.

\bibitem{Bauer:2000yr}
C.~W. Bauer, S.~Fleming, D.~Pirjol, and I.~W. Stewart,
\newblock Phys.Rev. {\bf D63}, 114020 (2001), hep-ph/0011336.

\bibitem{Idilbi:2008vm}
A.~Idilbi and A.~Majumder,
\newblock Phys.Rev. {\bf D80}, 054022 (2009), 0808.1087.

\bibitem{Ovanesyan:2011xy}
G.~Ovanesyan and I.~Vitev,
\newblock JHEP {\bf 1106}, 080 (2011), 1103.1074.

\bibitem{Ovanesyan:2011kn}
G.~Ovanesyan and I.~Vitev,
\newblock Phys.Lett. {\bf B706}, 371 (2012), 1109.5619.

\bibitem{Fickinger:2013xwa}
M.~Fickinger, G.~Ovanesyan, and I.~Vitev,
\newblock JHEP {\bf 1307}, 059 (2013), 1304.3497.

\bibitem{Altarelli:1977zs}
G.~Altarelli and G.~Parisi,
\newblock Nucl.Phys. {\bf B126}, 298 (1977).

\bibitem{Betz:2014cza}
B.~Betz and M.~Gyulassy,
\newblock (2014), 1404.6378.

\bibitem{Djordjevic:2013xoa}
M.~Djordjevic and M.~Djordjevic,
\newblock (2013), 1307.4098.

\bibitem{Wang:2009qb}
W.-t. Deng and X.-N. Wang,
\newblock Phys.Rev. {\bf C81}, 024902 (2010), 0910.3403.

\bibitem{Chang:2014fba}
N.-B. Chang, W.-T. Deng, and X.-N. Wang,
\newblock Phys.Rev. {\bf C89}, 034911 (2014), 1401.5109.

\bibitem{Gyulassy:2000fs}
M.~Gyulassy, P.~Levai, and I.~Vitev,
\newblock Phys.Rev.Lett. {\bf 85}, 5535 (2000), nucl-th/0005032.

\bibitem{Gyulassy:2000er}
M.~Gyulassy, P.~Levai, and I.~Vitev,
\newblock Nucl.Phys. {\bf B594}, 371 (2001), nucl-th/0006010.

\bibitem{Ellis:2010rwa}
S.~D. Ellis, C.~K. Vermilion, J.~R. Walsh, A.~Hornig, and C.~Lee,
\newblock JHEP {\bf 1011}, 101 (2010), 1001.0014.

\bibitem{Gallicchio:2011xq}
J.~Gallicchio and M.~D. Schwartz,
\newblock Phys.Rev.Lett. {\bf 107}, 172001 (2011), 1106.3076.

\end{thebibliography}

\end{document}